\documentclass[letterpaper]{article} % DO NOT CHANGE THIS
\usepackage[draft]{aaai25}  % DO NOT CHANGE THIS
\usepackage{times}  % DO NOT CHANGE THIS
\usepackage{helvet}  % DO NOT CHANGE THIS
\usepackage{courier}  % DO NOT CHANGE THIS
\usepackage[hyphens]{url}  % DO NOT CHANGE THIS
\usepackage{graphicx} % DO NOT CHANGE THIS
\usepackage{natbib}  % DO NOT CHANGE THIS AND DO NOT ADD ANY OPTIONS TO IT
\usepackage{caption} % DO NOT CHANGE THIS AND DO NOT ADD ANY OPTIONS TO IT
\usepackage{algorithm}
\usepackage{algorithmic}
\usepackage{amsmath}

\usepackage{xcolor}
\newcommand{\answerYes}[1]{\textcolor{blue}{#1}} 
\newcommand{\answerNo}[1]{\textcolor{teal}{#1}} 
\newcommand{\answerNA}[1]{\textcolor{gray}{#1}}

\usepackage{newfloat}
\usepackage{listings}
\DeclareCaptionStyle{ruled}{labelfont=normalfont,labelsep=colon,strut=off} % DO NOT CHANGE THIS
\floatstyle{ruled}
\newfloat{listing}{tb}{lst}{}
\floatname{listing}{Listing}
\usepackage{booktabs, censor}

\title{\censor{MurkySky}: Analyzing News Reliability on Bluesky}
\author{
    Vikas Reddy\textsuperscript{\rm 1,2},
    Giovanni Luca Ciampaglia\textsuperscript{\rm 2},
}
\affiliations{
    \textsuperscript{\rm 1}Johns Hopkins University, Baltimore, MD, USA\\
    \textsuperscript{\rm 2}University of Maryland, College Park, MD, USA\\
    vkasire2@jhu.edu, gciampag@umd.edu
}

\begin{document}

\StopCensoring

\maketitle

\begin{abstract}

Bluesky has recently emerged as a lively competitor to Twitter/X for a platform
for public discourse and news sharing. Most of the research on Bluesky so far
has focused on characterizing its adoption due to migration. There has been less
interest on characterizing the properties of Bluesky as a platform for news
sharing and discussion, and in particular the prevalence of unreliable
information on it. To fill this gap, this research provides the first
comprehensive analysis of news reliability on Bluesky. We introduce
\censor{MurkySky}, a public tool to track the prevalence of content from
unreliable news sources on Bluesky. Using firehose data from the summer of 2024,
we find that on Bluesky reliable-source news content is prevalent, and largely
originating from left-leaning sources. Content from unreliable news sources,
while accounting for a small fraction of all news-linking posts, tends to
originate from more partisan sources, but largely reflects the left-leaning skew
of the platform. Analysis of the language and hashtags used in news-linking
posts shows that unreliable-source content concentrates on specific topics of
discussion.  

\end{abstract}

% Uncomment the following to link to your code, datasets, an extended version or similar.
%
% \begin{links}
%     \link{Code}{https://aaai.org/example/code}
%     \link{Datasets}{https://aaai.org/example/datasets}
%     \link{Extended version}{https://aaai.org/example/extended-version}
% \end{links}

\section{Introduction}

Decentralization is a recent trend in social media that seeks to challenge the status quo offered by tech giants such as Facebook, Twitter/X, or TikTok. These novel networks offer a vision of user empowerment and distributed governance, shifting control away from large tech companies to individual users and collectives \cite{Zhang2024,hogg2024shaping}. Yet, despite their promise for greater agency and control, decentralized social media platforms also introduce challenges for researchers who wish to study them, and, given their relative novelty, also raise questions about the reliability of the information that is being shared on them \cite{Datta2010}.

One of the most interesting new players in this field is Bluesky: started initially within Twitter with the goal of creating a decentralized version of that platform, it later became an independent project with the goal, in direct competition with Twitter, of providing users with greater influence on content moderation decisions and more choice in algorithmic curation \cite{kleppmann2024bluesky}. It then became one of the beneficiaries of the `Great Twitter Migration' \cite{he2023flocking}, and has recently emerged as a serious competitor to Twitter/X. 

In part due to the circumstances behind its rise, much of the literature on Bluesky and other decentralized platforms such as Mastodon has so far focused on their adoption as a consequence of the aforementioned migration from Twitter/X~\cite{jeong2024exploring,lacava2021,he2023flocking,lacava2023drivers}. This is in part due to the fact that this migration offers the rare opportunity to observe in real time a process of platform--platform competition, reminiscent in some ways of the early rise of the previous generation of social networking services, such as Facebook and LinkedIn~\cite{ribeiro2015modeling,anderson2015global}.

However, less attention has been devoted to characterizing properties of these emerging platforms in and of themselves, not in relation to other legacy platforms. This is especially relevant, since the rise of decentralized services is in part in response to concerns about the degradation of civic discourse under the traditional (i.e. centralized) platform model \cite{doctorow2023social}, including concerns about the proliferation of inaccurate information on them \cite{shao2018spread,shao2018anatomy,lazer2018science}. It is thus important to assess whether Bluesky offers a viable alternative as a channel for reliable information. To address this gap, here we study the prevalence of unreliable news content on BlueSky, and characterize the audiences that share and engage with it.

In addition to this substantive gap, there are also methodological reasons that make it compelling to focus on Bluesky as a case study for decentralized social media beyond their mere user growth. Decentralization poses unique challenges to research. Under the decentralized model all data are, by design, distributed across different entities and organizations. In addition, on certain networks, like Mastodon, individual instance--nodes may actively discourage data collection on ethical and privacy grounds~\cite{waehner2024dont}. Thus it is important to develop novel data collection tools to facilitate independent research on these platforms.

To address this limitation, here we introduce \censor{MurkySky}, a public-facing tool designed to (\emph{i}) systematically collect public posts with news links shared on Bluesky, and (\emph{ii}) assess the prevalence of content from unreliable news source. To do this, \censor{MurkySky} relies on source reliability ratings as a coarse way to distinguish between reliable and unreliable sources, which have been shown to be robust across raters and assessment criteria~\cite{lin2023high}. Data from \censor{MurkySky} can be accessed via a REST API available at \censor{\url{https://rapidapi.com/csdl-umd-csdl-umd-default/api/murkysky-api}}.

Using \censor{MurkySky}, we collected a corpus of posts containing links to news sources and their associated reliability rating. We first aggregate posts based on their reliability, and tracked over time the prevalence of unreliable news content across the whole platform, observing temporal trends and fluctuations in the frequency of its posting and sharing by Bluesky users. We then extracted the contents of the posts (links, hashtags, terms, etc.) and performed a number of analyses to visualize the structure of news topics, and the major audiences that engage and share such posts. Finally, we turned to analyzing the news sources included in these posts. We identified news sources by extracting the Web domain information from the URLs of the links embedded in these posts. Focusing only on English-based sources, we then estimated the popularity and prevalence of news sources across the Left--Right political spectrum.  

Ultimately, by characterizing the prevalence, kind, and ideological composition of unreliable information on BlueSky, our study contributes to the growing literature on decentralized social media. Even though the results reported here provide only a snapshot in time of the evolution BlueSky, and may not generalize to other time periods, nonetheless \censor{MurkySky}  
provides a user-friendly way for social media researchers to continue monitoring the development of this platform over time. 

\section{Related Works}

Users migrate platforms in response to various \emph{push} and \emph{pull} factors. Push factors, such as dissatisfaction with platform policies or shifts in community environments, drive users away from their current platforms \cite{hou2020understanding}. For example, the acquisition of Twitter by Elon Musk led to widespread concerns about content moderation and platform policies, prompting many users to seek alternatives like Mastodon and Bluesky \cite{jeong2024exploring}. Pull factors, like the promise of enhanced privacy, decentralized control, and improved information quality attract users to new platforms that better align with their values \cite{jeong2024exploring}.

Of course, when users consider migrating to alternative social media platforms, their choices also reflect their priorities regarding community governance, technical infrastructure, and the balance between server autonomy and network-wide coordination \cite{devito2017platforms}. Understanding these distinct characteristics helps explain migration patterns. We can distinguish two major models, though both models distribute control away from central authorities \cite{raman2019}.

\emph{Federated networks} like Mastodon consist of independently managed servers, or \emph{instances}, that communicate through standardized protocols~\cite{lemmerwebber2018activitypub}. This setup promotes decentralization while ensuring coordination and consistency across servers \cite{lacava2021,lacava2023drivers}. Users benefit from increased control on their data, although many other aspects of their experience depend on the local instance policies, for example vis a vis privacy.

Instead, \emph{decentralized networks} like Bluesky emphasize a unified user experience similar to that of a legacy platform like Twitter/X, and relegate decentralization to its underlying protocol~\cite{kleppmann2024bluesky}. This model still addresses some of the fundamental issues of data ownership and platform governance at the heart of user dissatisfaction with centralized platforms~\cite{quelle2024}, while facilitating migration through the deployment of a more familiar user experience.  

While decentralization can reduce the information asymmetry between users and platform operators that is motivating migration from centralized systems, it is important to note that it does not automatically preserves user privacy -- a key consideration for migrating users. In fact, in the current implementation of Bluesky, any relay operators can access user data, including direct messages and block lists. Nonetheless, the approach of Bluesky to decentralizing governance may be attracting users who are specifically concerned about censorship and algorithmic manipulation in traditional systems, which may also influence the kind of content and sources they choose to share and engage with on the platform~\cite{goel2010anatomy}.  

Finally, the quality of information and the health of civic discussions plays a crucial role in platform dynamics and user migration patterns. While decentralized platforms may initially attract users seeking reliable information, their growth can paradoxically make them more attractive targets for actors seeking to spread misinformation \cite{cohen2020, acemoglu2010}. This highlights the importance of developing measurement tools to monitor and understand the prevalence of unreliable content on these platforms~\cite{Shao2016}. These tools are essential not only for platform operators and researchers, but also for ordinary users, who seek to make informed decisions about which platforms to join and trust. One such tool is the Iffy Quotient, which our \censor{MurkySky} tool is inspired to, and which tracks the share of content from `iffy' (i.e. unreliable) sources on Twitter and Facebook~\cite{resnick2023iffy}.

Of course, such tools require a technical infrastructure for the collection and analysis of data from these platforms. Researchers have traditionally relied on tools for programmatic access, like the Pushshift API or the Twitter API to analyze centralized social media platforms like Reddit or Twitter, respectively
\cite{baumgartner2020pushshift, murtfeldt2024rip}. Unlike other decentralized platforms such as Mastodon, where data access is fragmented across instances, and thus there is not a single entry point capable of providing access to data about the entire network, the Bluesky protocol includes a firehose primitive that gives access to the full stream of messages on the platform \cite{schneider2019decentralization}. 

\section{Methodology}
\label{sec:methods}

This section briefly describes the design of \censor{MurkySky}, as well as our approach for studying how news content is shared and engaged with on Bluesky. We briefly give details in particular on hashtag co-occurrence networks, audience segmentation, the political orientations of news sources, and rank--frequency distributions to uncover media consumption trends.

\subsection{The \censor{MurkySky} Tool}
\censor{MurkySky} collects and analyzes links shared on the Bluesky platform by connecting to the Firehose API, a primitive in the AT protocol that provides a real-time feed of user activities such as posts, likes, follows, and handle changes. Each Personal Data Server (PDS) is responsible for managing a stream of activity for its assigned repositories, and when data is requested, relays aggregate the relevant streams from multiple PDSs into a unified feed, effectively creating a comprehensive firehose of user interactions (cf. \url{https://docs.bsky.app/docs/advanced-guides/firehose}).

\censor{MurkySky} captures all these events (i.e. posts, likes, and reposts), extracting any embedded URIs included in them. When a user reposts or likes content, their Personal Data Server (PDS) stores a strong reference to the original content in another PDS. To retrieve the original post, \censor{MurkySky} utilizes the \texttt{app.bsky.feed.getPosts} endpoint from the Bluesky Lexicon APIs, fetching the record from the PDS that originally created the post. The extracted URLs are then evaluated using the NewsGuard rating system (see below) to assess the reliability of the information from the news sources, if any is present in the post.

The link sharing and reliability data processed by \censor{MurkySky} is visualized using Shiny for Python, with two main approaches to presenting news source reliability patterns: `Relative' and `Absolute' measurements. The `Relative' view displays the ratio of unreliable to reliable news source links shared on Bluesky over various time frames -- 1 week, 30 days, the entire dataset, or a custom period -- providing insights into how the proportion of reliable versus unreliable news sources changes over time. In contrast, the `Absolute' view offers a breakdown of these quantities by showing three separate trend lines: the total volume of news links shared, the number of links from reliable news sources, and the number of links from unreliable sources according to NewsGuard ratings. These visualization options are available in both hourly and daily aggregations, enabling a comprehensive analysis of both the volume and reliability trends of news sources being shared across the Bluesky platform.

\subsection{Source Reliability Assessment}

To determine the prevalence of unreliable content, \censor{MurkySky} relies on source reliability ratings provided by NewsGuard. The NewsGuard rating system provides a systematic approach to evaluating news source credibility. Their methodology assesses news sources based on nine non-partisan, impartial criteria focused on credibility and transparency, such as avoiding false or misleading content, responsibly reporting and presenting information, and clearly identifying advertising \cite{newsguard2024}. Trained journalists evaluate each news source using this standard rubric and, if any of the criteria was unmet, reach out to the editorial team at the news source to offer them the chance of a response. Senior editors then review and fact-check each rating to ensure fairness and accuracy. Ratings are regularly updated to reflect changes in editorial practices of a source. News sources that score at or above 60 on the rubric adhere to basic standards of credibility and transparency and are rated reliable; scores below 60 typically indicate that a news source does not meet basic journalistic standards, and is thus rated unreliable \cite{bianchi2024evaluating}.

\subsection{Hashtag Co-occurrence Analysis}
This study explores hashtag co-occurrence in posts containing news links on the Bluesky platform to analyze discussions around trustworthy and untrustworthy news sources. An undirected graph is constructed where nodes represent hashtags; in this graph there is an edge between two hashtags if they co-occur within one or more posts. The weight of each edge is a measure of the reliability of the posts in which two given hashtags appear together. This is calculated by considering all posts with the two co-occurring hashtags that have a link to some news source, and computing the difference between the frequency of reliable-source link-posts to that of unreliable-source ones, normalized by the total frequency link-posts with the two hashtags, that is
\begin{equation}
\text{Edge Weight} = \frac{W_{\rm UT} - W_{\rm T}}{W_{\rm UT} + W_{\rm T}}
\label{eq:weight}
\end{equation}

\noindent where \( W_{\rm UT} \) (respectively \(W_{\rm T}\)) refers to the number of posts containing NewsGuard-rated links with scores below 60 (resp. at or above 60) where both hashtags co-occur. The weight of each node is then calculated based on the average weight of all its incident edges. This average reflects the overall trustworthiness of the hashtag in relation to its other co-occurring hashtags. The resulting graph, with weighted nodes and edges, is visualized using the network analysis tool Gephi.

\subsection{Audience Segmentation}

We also performed a $k$-core analysis of the network of likes and reposts on Bluesky. This network, where each user is a node and there is an undirected edge between two users if one of the two liked or reposted content from the other one, provides fine-grained information into the types of audiences that are responsible for the dissemination of news content on the platform. By identifying the highest $k$-core value \( k = 38 \), we focused on the most active users who repost and like posts containing news links, and then segmented them into distinct modularity classes based on their interaction patterns and content preferences. 

To better characterize these audience segments, we created a word cloud to visualize the most prominent terms associated with these highly active user groups. The word cloud was constructed by analyzing the `user timeline' of each user -- the feed comprising of posts and reposts made by the user -- within each modularity class, creating a bag of words for the text corpus of each class, as well as a total bag of words combining the corpora from all the modularity classes. Log-odds ratios with informative Dirichlet priors were then applied to identify terms that are significantly more likely to appear in posts from each modularity class compared to the whole corpus overall \cite{Monroe2017}. The log-odds ratio for a word \( w \), denoted as \( \delta^{(i-j)}_w \), is estimated as:
\begin{multline}
\delta^{(i-j)}_w = \log \left( \frac{y_{iw} + a_w}{n_i + a_0 - y_{iw} + a_w} \right) - \\ \log \left(\frac{y_{jw} + a_w}{n_j + a_0 - y_{jw} + a_w} \right)
\label{eq:logodds}
\end{multline}

\noindent where \( y_{iw} \) is the count of word \( w \) in the target modularity class, \( y_{jw} \) is the combined count of word \( w \) in the other modularity classes, \( a_w \) is the total count of word \( w \) across all modularity classes, \( n_i \) is the total count of all words in the target modularity class, \( n_j \) is the total count of all words in the other modularity classes, and \( a_0 \) is the total count of all words across all modularity classes. 

This formula measures how much more likely a word is to appear in the target modularity class compared to the other modularity classes, effectively highlighting distinctive terms and themes. By comparing the relative frequencies of words across different classes, the log-odds ratio reveals key topics and vocabulary that are characteristic of the interactions and interests of each class, providing insights into the content preferences and engagement patterns of users within each modularity class.

\subsection{Media Source Political Orientation}
To classify the political orientation of media sources, we combined three datasets -- Media Bias Fact Check (\url{https://mediabiasfactcheck.com/}), AllSides (\url{https://www.allsides.com/unbiased-balanced-news}), and NewsGuard (\url{https://www.newsguardtech.com/}) -- providing a balanced, and high-coverage analysis of political leanings and reliability for news shared on the Bluesky platform. The process began with data collection using Media Bias Fact Check Data Collection, a Python script developed by the Center for Computational Analysis of Social and Organizational Systems at CMU to compile a CSV file containing the political orientation of media outlets (\url{https://github.com/CASOS-IDeaS-CMU/media_bias_fact_check_data_collection}). This tool provided a strong starting point by offering bias ratings for over $2,000$ media outlets. 

To enhance the comprehensiveness and reliability of our dataset, we developed a hierarchical approach to combining multiple rating systems. We prioritized Media Bias Fact Check as our primary source due to its higher credibility and detailed analysis of news sources. For outlets not covered by Media Bias Fact Check, we turned to AllSides ratings, as both systems offer granular political bias classifications (such as ``Left-Leaning'' versus simply ``Left''). Finally, for sources not covered by either Media Bias Fact Check or AllSides, we incorporated NewsGuard ratings, which helped identify additional unreliable sources and complete our coverage. During the consolidation process, we adhered to this strict hierarchy to resolve any rating discrepancies between systems: Media Bias Fact Check ratings took precedence, followed by AllSides, and then NewsGuard. This systematic approach allowed us to develop a unified dataset that comprehensively combines political bias and trustworthiness assessments into a single CSV file, maximizing coverage while maintaining consistent classification standards.

After collecting all NewsGuard links from June to August, 2024, we categorized media sources based on their ratings. We relied on the threshold suggested by NewsGuard (see above). This process let us generate a chart illustrating the distribution of political orientations across reliable and unreliable news links shared on the Bluesky platform. The final visualization revealed patterns in media consumption and reliability, offering insights into the political leanings of the shared news links.

\begin{figure*}[t]
    \centering
    \includegraphics[width=0.495\textwidth]{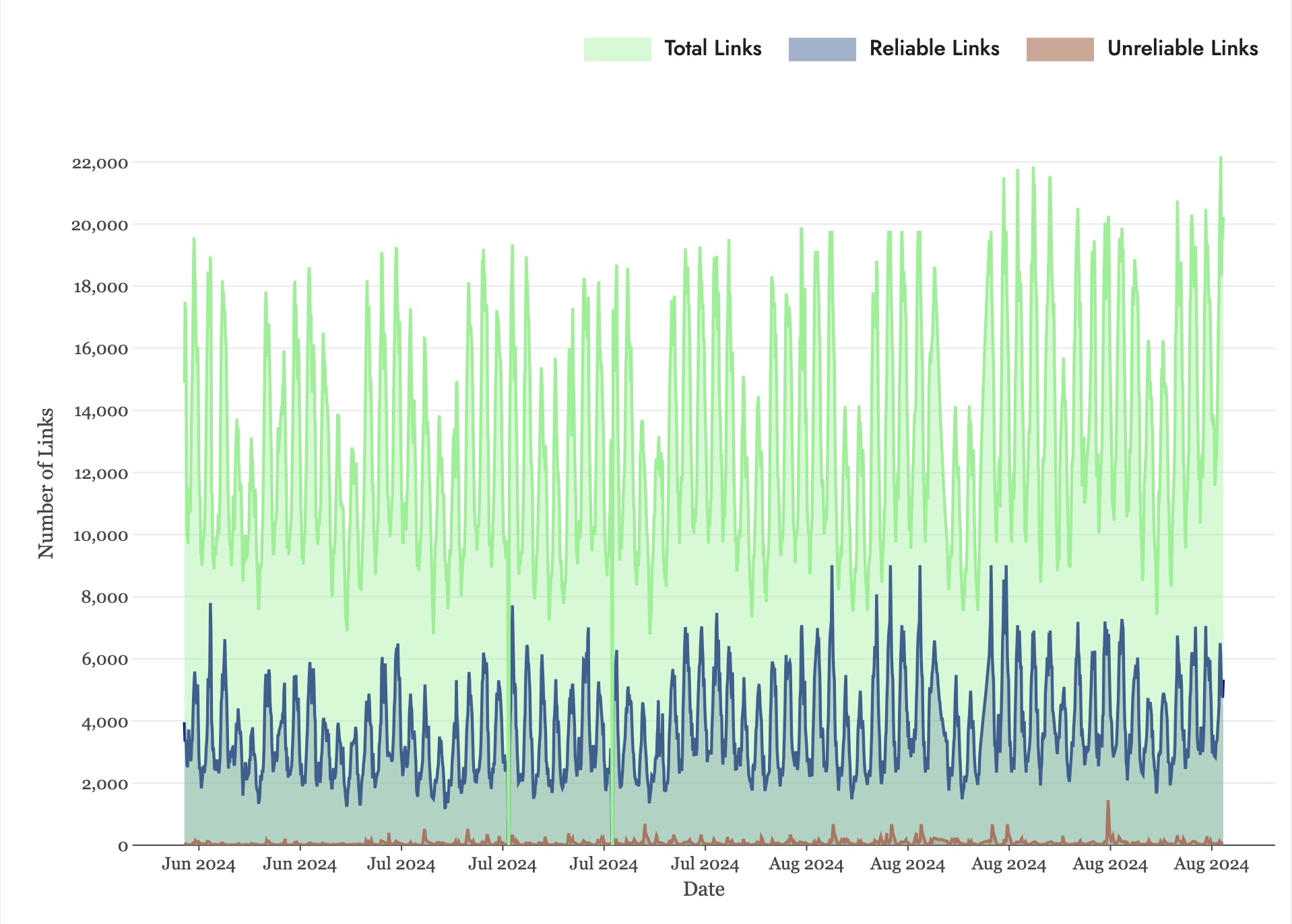}
    \includegraphics[width=0.495\textwidth]{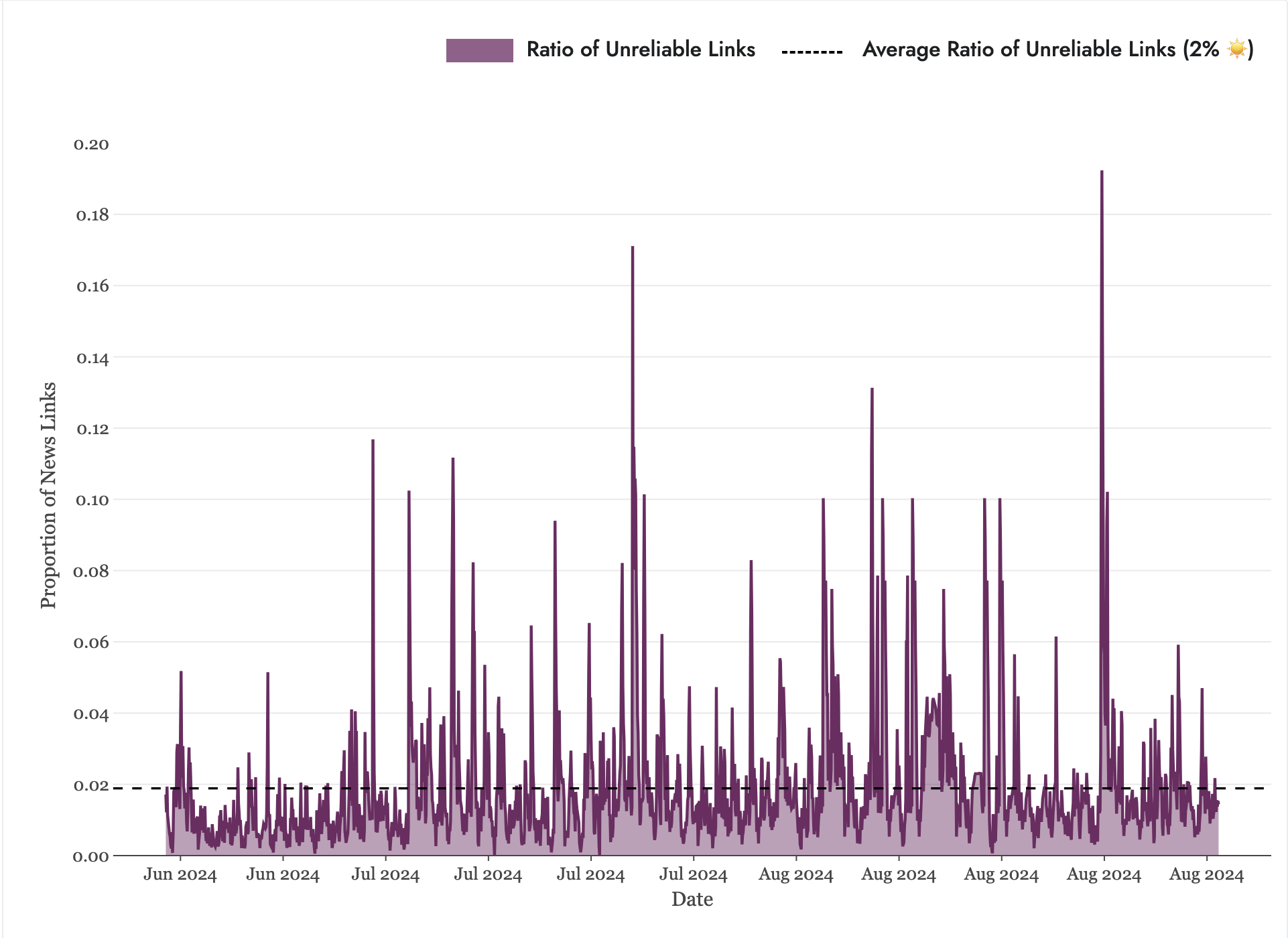}
    \caption{Left: Hourly total counts of reliable, unreliable, and total news links on Bluesky. Right: Proportion of unreliable news links relative to total news links on Bluesky.}
    \label{fig:figure_1_absolute_relative}
\end{figure*}

\subsection{Rank--Frequency of NewsGuard Domains}

To investigate the rank--frequency distribution of media sources on Bluesky, we applied statistical analysis to identify patterns of media source prominence and compare distributions between reliable and unreliable sources as classified by NewsGuard.

We began by extracting and preprocessing links shared on Bluesky between June and August, 2024, isolating their domains using base URLs to ensure consistency and eliminate duplicates. These domains were then classified as `reliable' or `unreliable' based on NewsGuard ratings (see above), resulting in two separate datasets. Next, we performed a rank--frequency analysis for each dataset. Domains were grouped by their frequency of occurrence, and were ranked in descending order. This approach aligns with heavy-tailed distributions, which describe how a small number of elements often dominate occurrences in systems ranging from language to information networks \cite{cristelli2012zipf}.

To analyze the distributions, we plotted rank against frequency for both reliable and unreliable domains. Given the heavy-tailed characteristics typically observed in rank-frequency distributions, we employed a logarithmic scale for both axes (log--log plots). This transformation allowed us to visualize the steepness of frequency declines and compare the distributional slopes of reliable versus unreliable domains. A sharper decline for reliable domains would indicate greater concentration among a few dominant sources, while a flatter curve for unreliable domains would suggest a more diffused, less hierarchical distribution.

Lastly, we performed a combined rank--frequency analysis, merging reliable and unreliable domains into a single dataset. This allowed us to observe overarching patterns of source dominance across the entire platform and identify the extent to which reliable and unreliable sources coexisted within the top ranks.

\section{Results}
\subsection{Deployment of \censor{MurkySky}}

We deployed \censor{MurkySky} on December 8th, 2023. After an initial phase during which only posts, but not their reposts, were collected, we started collecting reposts too on June 14th, 2024. Here we report results for the period from June 14th, 2024 to August 29, 2024, a period of slow, but steady growth, which occurred prior to the major influx of users that occurred in the wake of the 2024 US Presidential Elections. These data reveal that unreliable information on Bluesky did not occur frequently at the time, contributing only a small portion of the total news content available on the platform. 

The Absolute Values chart (left panel of Fig.~\ref{fig:figure_1_absolute_relative}), which tracks the total number of links categorized by NewsGuard ratings each day, shows that unreliable news links were relatively scarce during the observation period. The chart reveals a consistent daily pattern, with the number of news articles shared peaking in the afternoon Greenwich Mean Time (or morning Eastern Standard Time) and decreasing throughout the rest of the day~\cite{kates2021times}.

The Relative Values chart (right panel of Fig.~\ref{fig:figure_1_absolute_relative}), which examines the proportion of unreliable links compared to the total number of links with a NewsGuard rating, further confirms the rarity of content from unreliable sources. Although there are occasional spikes in the ratio of unreliable links with a NewsGuard rating, these instances are infrequent. On average, only about 2\% of the links with a NewsGuard rating are from an unreliable source. This suggests that while there are brief periods of increased sharing of unreliable source, the overall proportion of their content remained very low throughout the summer of 2024.

\begin{figure*}[t]
    \centering
    \includegraphics[width=0.9\linewidth]{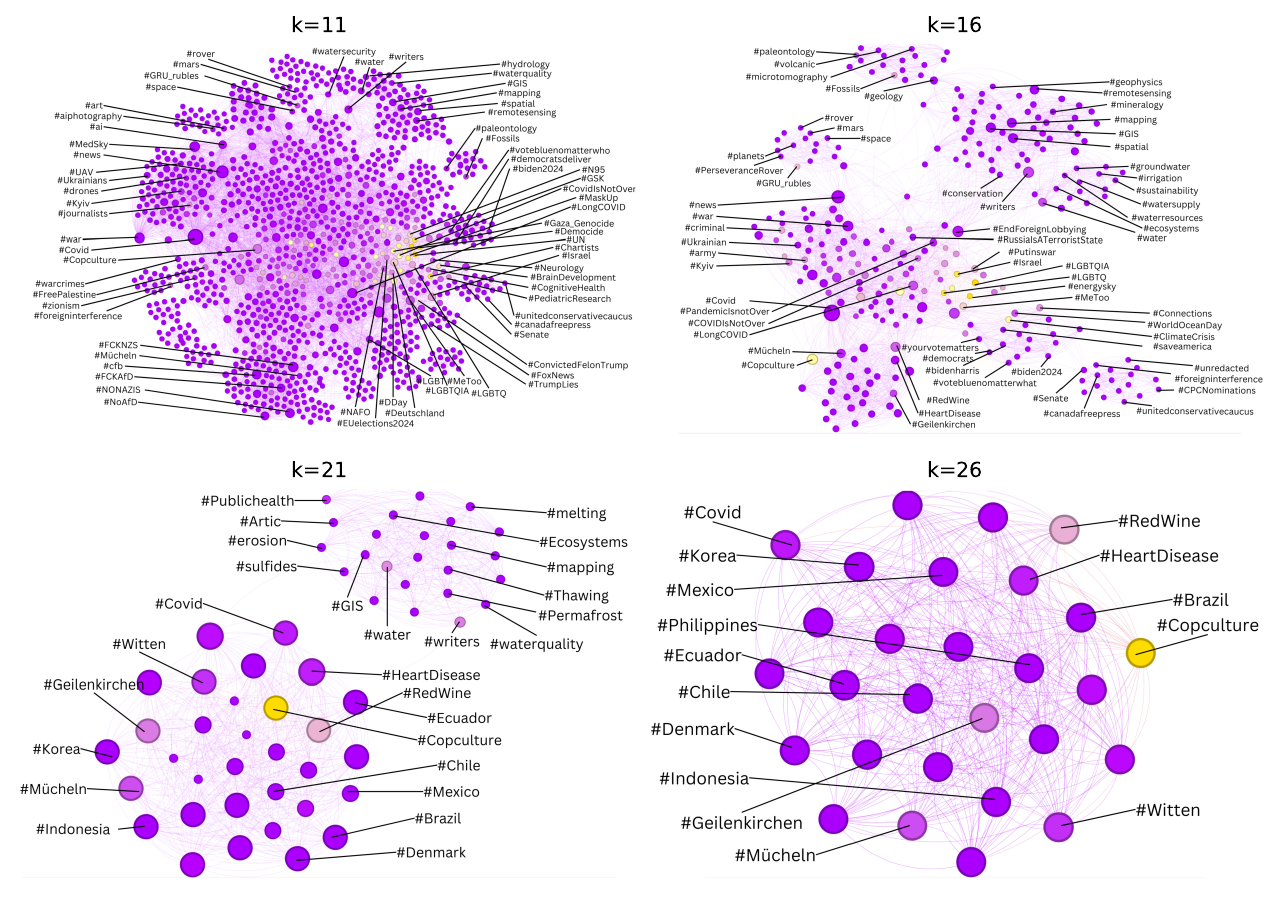}
    \caption{Hashtag co-occurrence for different $k$-core networks ($k=11, 16, 21, 26$). Node color indicates the reliability of news links shared with each hashtag (Yellow = unreliable, Purple = Reliable). Node position computed with a force-directed layout.}
    \label{fig:figure_3_hashtag_cooccurance}
\end{figure*}

\subsection{Hashtag Co-occurrence Analysis}

We then extract $k$-core hashtag co-occurrence networks for different values of
$k$ (see Fig.~\ref{fig:figure_3_hashtag_cooccurance}). Combined with our
weighting scheme (see Eq.~\ref{eq:weight}), the hashtags with which Bluesky
users choose to describe their posts provide a rough indication of what topics
were associated to news content, and the reliability of their sources. Overall,
and consistent with the fact that unreliable sources were rare on Bluesky during
the observation period, most hashtags appeared exclusively in conjunction with
content from reliable sources. However, unreliable-source content tended to be
concentrated around certain topics only, and not uniformly distributed across
topics. By producing different network maps for different values of $k$, we can
thus evaluate the centrality of unreliable-source content within the broader
semantic structure of Bluesky conversations.

At the maximum core value ($k=26$), the network predominantly captures global topics and significant events. Prominent hashtags include \texttt{\#Covid}, along with country-specific tags like \texttt{\#NewZealand}, \texttt{\#Korea}, \texttt{\#Mexico}, \texttt{\#Brazil}, and \texttt{\#Philippines}. At this core level, content from unreliable news sources is rare and primarily clustered around two distinct topics: \texttt{\#CopCulture} and a set of German location-based hashtags (\texttt{\#Witten}, \texttt{\#Mücheln}, and \texttt{\#Geilenkirchen}). These German hashtags represent cities where massive demonstrations occurred against the AfD (Alternative for Germany) party following revelations about a secret meeting where party members and other far-right figures allegedly discussed plans for ``remigration'' -- a dog-whistle for the mass deportation of immigrants and German citizens with immigrant backgrounds~\cite{mendelsohn2023dogwhistles}, drawing widespread comparisons to Nazi-era policies. 

For $k=21$, the network maintains the global focus observed before, but
introduces an additional cluster centered on environmental and scientific
topics. This includes hashtags like \texttt{\#PublicHealth}, \texttt{\#Spatial},
\texttt{\#Ecosystems}, \texttt{\#GIS}, \texttt{\#Hydrology}, and
\texttt{\#Melting}. These hashtags suggest an emphasis on climate change and
geographical research. Content from unreliable sources is associated to hashtags
such as \texttt{\#Water} and \texttt{\#Writers}, hinting at possible
controversies or misrepresentations related to environmental and public health
issues.

For $k=16$, we see a more diversified network, reflecting a broader array of topics. In addition to the environmental and geographical clusters, the network encompasses discussions on political figures like Biden and Trump, the climate crisis, planet exploration, and paleontology. Content from unreliable news sources becomes notably concentrated around sensitive subjects including LGBTQIA+, the Israel--Hamas conflict, the MeToo movement, climate crisis, and the aforementioned \texttt{\#CopCulture}. All these involve contentious social, political, and environmental issues.

Finally, for $k=11$ we see a network where unreliable-source content is primarily on global health and geopolitical conflicts. Key topics include Covid-19, the U.N., eugenics, the aforementioned Israel--Hamas conflict, and neurology. 

\subsection{Bluesky News Audience Segmentation}

\begin{figure}[ht]
    \centering
    \includegraphics[width=1.0\linewidth]{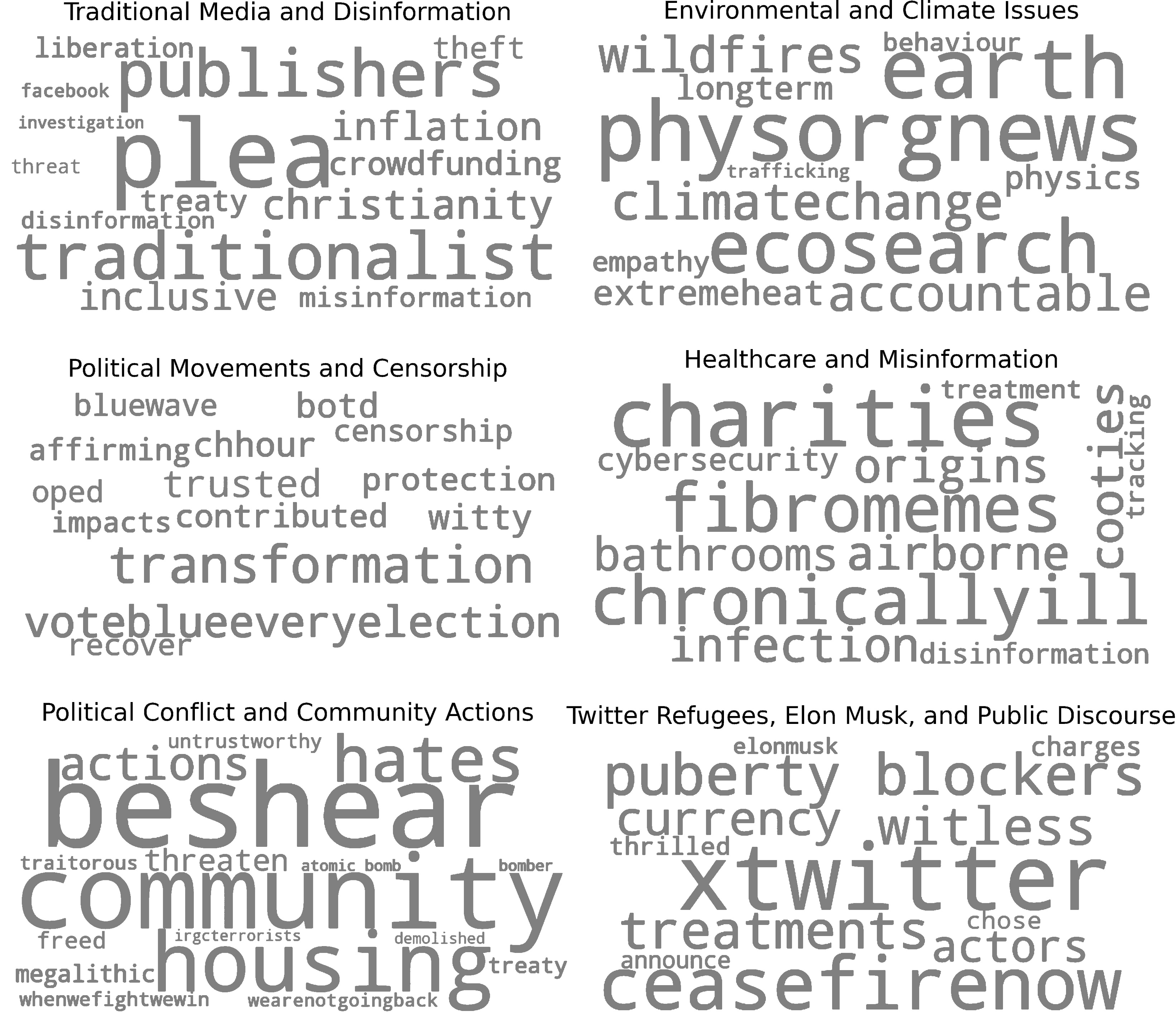}
    \caption{Word clouds for each modularity class from the max-$k$-core of the likes and repost network for posts with links to news sources. The size of each word is proportional to its log-odds ratio (see Eq.~\ref{eq:logodds}).}
    \label{fig:figure_4_word_cloud}
\end{figure}

To better understand what type of audiences were engaged with news content
during the study period, we extracted unigrams from the `user timeline' feeds of users, focusing
in particular on their posts and reposts of news-linking content. We then
defined a network of likes and reposts using these reposts and likes, and
applied community detection to segment the network in distinct audiences of
users \cite{Blondel_2008}, each associated with a specific corpus of unigrams describing their
reposts and like. We then used Eq.~\ref{eq:logodds} to compute the log-odds of
each unigram across the various corpora, and visualize word clouds for the most
representative words in each corpus, see Fig.~\ref{fig:figure_4_word_cloud}. The
analysis of these community-drawn corpora, which were drawn only from users in
the max-$k$-core of the underlying network ($k=38$), reveals six prominent
groups, each with distinct thematic focuses. These groups reflect the varied
interests and priorities of Bluesky users, especially in relation to
news reliability and content dissemination.

\textbf{Traditional Media and Disinformation}: This group is characterized by terms associated with both traditional media practices and the spread of misinformation. Key terms include ``liberation,'' ``publishers,'' and ``investigation,'' which highlight a focus on media operations and integrity. Additionally, words like ``disinformation'' and ``misinformation'' highlight the group's engagement with issues of media credibility and the challenges posed by misleading information. The presence of terms such as ``facebook,'' ``inflation,'' and ``crowdfunding'' suggests an interest in how traditional media intersects with broader societal and economic issues.

\textbf{Environmental and Climate Issues}: Users in this group focus on environmental and climate-related topics, which can be seen from terms such as ``wildfires,'' ``climatechange,'' and ``extremeheat.'' These words indicate a concern for long-term environmental impacts and behaviors affecting the planet. The inclusion of ``empathy'' and ``accountable'' reflects a call for responsible and considerate action towards environmental issues, with ``physorgnews'' and ``ecosearch'' pointing to sources and platforms that are engaged in scientific discourse about climate change.

\textbf{Political Movements and Censorship}: This group focuses on political activism, particularly from the perspective of the U.S. Democratic party, as indicated by terms like ``bluewave'' and ``voteblueeveryelection.'' The presence of ``censorship'' points to concerns about freedom of speech and political influence. The term ``oped'' refers to opinion pieces from news sources like \emph{The New York Times}, reflecting the group's interest in political commentary. Additionally, ``transformation'' suggests a focus on political change and its impact on public discourse. 

\textbf{Healthcare and Misinformation}: This group addresses healthcare related topics and misinformation within the medical field. Terms such as ``treatment,'' ``tracking,'' and ``fibromemes'' suggest a focus on health conditions, including chronic illnesses like fibromyalgia, and the spread of misleading or humorous content about them. Words like ``disinformation'' and ``chronicallyill'' point to concerns about how unreliable content can affect health perceptions and treatment options. The group's attention to ``airborne'' and ``infection'' highlights ongoing issues related to disease spread and public health communication.

\textbf{Political Conflict and Community Actions}: The language in this group reflects a focus on political conflict and community responses, with a particular emphasis on U.S. Democratic perspectives. Terms like ``actions,'' ``untrustworthy,'' and ``traitorous'' suggest concerns with political disputes and perceptions of betrayal or mistrust. The inclusion of slogans related to the 2024 U.S. Presidential Election, such as ``whenwefightwewin'' and ``wearenotgoingback'' indicates discussions on how Kamala Harris and her party could address and resolve global issues.

\textbf{Twitter Refugees, Elon Musk, and Public Discourse}: This group consists of users who migrated from Twitter and seek a similar social media experience but are dissatisfied with Elon Musk's actions and the changes implemented under his leadership. Terms like ``xtwitter,'' ``elonmusk,'' and ``witless'' reflect their frustration with the transformation of the platform. Additionally, the focus on ``puberty blockers'' and ``currency'' highlights ongoing debates around social issues and their impact on public discourse. This vocabulary demonstrates desire for a more open and user-centered experience, contrasting with the changes and perceived shortcomings on Twitter.

\begin{figure*}
    \centering
    \includegraphics[width=0.495\textwidth]{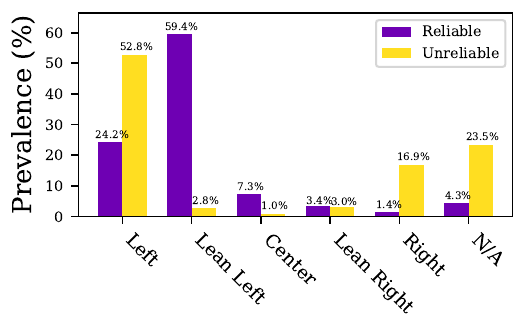}
    \includegraphics[width=0.495\textwidth]{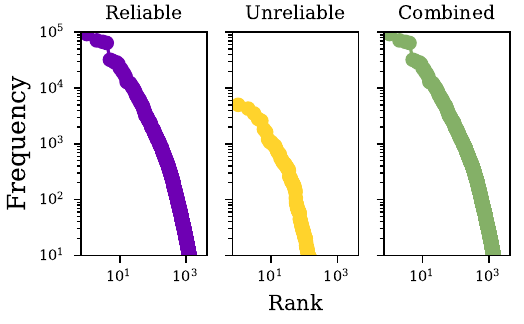}
    \caption{Left: Content prevalence, by source orientation and reliability. Right: Rank--Frequency plots of news source popularity}
    \label{fig:figure_5_orientation_6_rank_frequency}
\end{figure*}

\subsection{News Source Political Orientation}

Then, we plotted the prevalence of news-source content, focusing on the relationship between source political orientation and credibility, see Fig.~\ref{fig:figure_5_orientation_6_rank_frequency} (Left). For this analysis, we considered only English-language sources. Consistent with the audience segmentation findings, we observe a prevalence of left or left-leaning sources. Specifically, among reliable sources, 59.44\% are classified as `Lean Left,' with an additional 24.18\% positioned on the `Left,' while only 3.4\% and 16.9\% are either `Lean Right' or `Right', respectively. This pattern applies to unreliable sources too, though unreliable sources tend to be more partisan (i.e. less leaning in their orientation), compared to reliable ones. While 52.81\% of unreliable sources are categorized as 'Left,' and 16.95\% are classified as 'Right,' only 2.8\% and 3.0\% are `Lean Left' and `Lean Right', respectively.

\subsection{Rank--Frequency of NewsGuard Domains}

The right panel of Fig.~\ref{fig:figure_5_orientation_6_rank_frequency} shows the rank--frequency plot for the popularity of news sources on Bluesky. Consistent with much of the literature, we observe a tendency for few popular news sources to dominate the popularity rankings in both categories. For reliable sources, the top entries (see Table~\ref{tab:top-domains}) correspond to established news outlets like \emph{The New York Times} and \emph{The Guardian}. For unreliable sources, the top entries include a mix of national news outlets (e.g. \emph{MSNBC}, \emph{Daily Kos}), emerging local news networks (e.g. \emph{ohiocapitaljournal.com}), and sources run by foreign entities (e.g. \emph{censor.net}, \emph{globaltimes.cn}). 

\begin{table}
  \centering
  \caption[Top 10 Reliable and Unreliable Domains]{Top 10 Reliable and Unreliable Domains}
  \small
  \begin{tabular}{cll}
    \toprule
    \textbf{Rank} & \textbf{Reliable Domains} & \textbf{Unreliable Domains} \\
    \midrule
    1 & theguardian.com & dailykos.com \\
    2 & nytimes.com & msnbc.com \\
    3 & bbc.com  & thegatewaypundit.com \\
    4 & washingtonpost.com & wsws.org \\
    5 & spiegel.de & democracydocket.com \\
    6 & cnn.com & ohiocapitaljournal.com \\
    7 & reuters.com & middleeastmonitor.com \\
    8 & nbcnews.com & trtworld.com \\
    9 & npr.org & newsfromthestates.com \\
    10 & rawstory.com & globaltimes.cn \\
    \bottomrule
  \end{tabular}
  \label{tab:top-domains}
\end{table}

This disparity could perhaps explain the reason why content from unreliable sources was still rare on Bluesky during the summer of 2024, and is likely a reflection of the particular audiences present on the platform, which tended to skew on the left and favor more established (and thus also more reliable) news outlets~\cite{lazer2018science}.

\section{Limitations and Ethical Considerations}

This study offers valuable insights into news source reliability and user behavior on Bluesky, but several limitations should be acknowledged. One significant limitation is related to the real-time nature of the data collection process. \censor{MurkySky} uses a WebSocket connection to collect data about events (posts, reposts, likes, etc.) from the Bluesky Firehose. Because the Bluesky Firehose delivers only references to reposts, but not their complete payload, collection of reposts required additional processing time, which at times caused our script to lag behind the stream of events and disconnect from the stream. To minimize potential bias from data collection outages, here we reported results from a restricted observation window, from June to August 2024, during which our script ran without interruptions. This means our findings are only a snapshot in time and do not represent the full period of activity of Bluesky since its inception. In particular, variations in hashtag usage and language within this time frame can impact the generalizability of the co-occurrence relationships and the identification of significant terms in both the hashtag co-occurrence and audience segmentation analyses. 

The assessment of news source reliability using NewsGuard ratings also has its limitations. While NewsGuard offers a comprehensive evaluation rubric based on independent, non-partisan criteria, it relies on a mix of manual annotation and social listening to identify news sources to rate. This is slow a process and cannot identify each and every news source, especially those in the long tail of the popularity distribution, like niche or emerging news outlets~\cite{reuben2024leveraging}. Furthermore, focusing our reliance on source-level ratings as opposed to content-level ones necessarily means our prevalence measurements are coarse-grained estimates of the true prevalence of unreliable content on the platform~\cite{green2024curation}. Indeed, unreliable sources are known to often republish content from reliable sources (for example as part of syndication agreements)~\cite{shao2018spread}. In this sense, our results should be seen as a likely upper bound on the true prevalence of unreliable content.

Ethically, we follow best practices from the literature on public social media. Our social listening tool (\censor{MurkySky}) is designed to collect public data only, and even though we do not keep track of post deletions, and thus cannot honor user intentions about content deletions, all our results are reported in the aggregate only and do not include disaggregated information about individual users or posts. Likewise, our content analyses focus on hashtags and news sources, and do not mention individual users or posts.  

\section{Conclusion}

This study presents a comprehensive analysis of news source reliability and user behavior on Bluesky, revealing that unreliable information constitutes only about 2\% of the total news content on the platform, suggesting a relatively high standard of information reliability. The hashtag co-occurrence analysis uncovers distinct patterns of misinformation, particularly around sensitive topics such as political conflicts and environmental issues, while the $k$-core word cloud analysis highlights the diverse thematic focuses among user groups, ranging from environmental concerns to political activism. 

We find that on Bluesky reliable content is prevalent and left-leaning, with more than 90\% of news content that was posted or reposted on Bluesky originating from reliable sources (according to NewsGuard) and, of these, more than 80\% of news-posts and reposts originating from left or left-leaning sources. In contrast, when it come to (much rarer) unreliable-source content, the fraction of said content originating from right or right-leaning sources is somewhat higher, but still dominated by a prevalence of content from left or left-leaning sources.   

As Bluesky grows in popularity, and more users join it, either by migrating from different platforms or as their first social media platform, it is not clear whether these trends will persist, or whether the prevalence of unreliable-source content will inevitably increase. We hope that our research tool, \censor{MurkySky}, can help Bluesky users make sense of future trends and serve as a tool for technologists and policy makers to reflect on the health of this novel social media platform. 

%\section{Acknowledgements}
%
%G.L.C. is partially support by the National Science Foundation under award no.~2239194.

\bibliography{aaai25}

\begin{thebibliography}{40}
\providecommand{\natexlab}[1]{#1}

\bibitem[{Acemoglu, Ozdaglar, and ParandehGheibi(2010)}]{acemoglu2010}
Acemoglu, D.; Ozdaglar, A.; and ParandehGheibi, A. 2010.
\newblock Spread of (mis)information in social networks.
\newblock \emph{Games and Economic Behavior}, 70(2): 194--227.

\bibitem[{Anderson et~al.(2015)Anderson, Huttenlocher, Kleinberg, Leskovec, and Tiwari}]{anderson2015global}
Anderson, A.; Huttenlocher, D.; Kleinberg, J.; Leskovec, J.; and Tiwari, M. 2015.
\newblock Global Diffusion via Cascading Invitations: Structure, Growth, and Homophily.
\newblock In \emph{Proceedings of the 24th International Conference on World Wide Web}, WWW '15, 66–76. Republic and Canton of Geneva, CHE: International World Wide Web Conferences Steering Committee.
\newblock ISBN 9781450334693.

\bibitem[{Baumgartner et~al.(2020)Baumgartner, Zannettou, Keegan, Squire, and Blackburn}]{baumgartner2020pushshift}
Baumgartner, J.; Zannettou, S.; Keegan, B.; Squire, M.; and Blackburn, J. 2020.
\newblock The Pushshift Reddit Dataset.
\newblock In \emph{Proceedings of the International AAAI Conference on Web and Social Media}, volume~14, 830--839.

\bibitem[{Bianchi et~al.(2024)Bianchi, Pratelli, Petrocchi, and Pinelli}]{bianchi2024evaluating}
Bianchi, J.; Pratelli, M.; Petrocchi, M.; and Pinelli, F. 2024.
\newblock Evaluating Trustworthiness of Online News Publishers via Article Classification.
\newblock In \emph{Proceedings of the 39th ACM/SIGAPP Symposium on Applied Computing}, SAC '24, 671--678. New York, NY, USA: Association for Computing Machinery.
\newblock ISBN 9798400702433.

\bibitem[{Blondel et~al.(2008)Blondel, Guillaume, Lambiotte, and Lefebvre}]{Blondel_2008}
Blondel, V.~D.; Guillaume, J.-L.; Lambiotte, R.; and Lefebvre, E. 2008.
\newblock Fast unfolding of communities in large networks.
\newblock \emph{Journal of Statistical Mechanics: Theory and Experiment}, 2008(10): P10008.

\bibitem[{Cohen et~al.(2020)Cohen, Moffatt, Ghenai, Yang, Corwin, Lin, Zhao, Ji, Parmentier, P’ng, Tan, and Gray}]{cohen2020}
Cohen, R.; Moffatt, K.; Ghenai, A.; Yang, A.; Corwin, M.; Lin, G.; Zhao, R.; Ji, Y.; Parmentier, A.; P’ng, J.; Tan, W.; and Gray, L. 2020.
\newblock Addressing misinformation in online social networks: Diverse platforms and the potential of multiagent trust modeling.
\newblock \emph{Information (Switzerland)}, 11(11): 1--40.

\bibitem[{Cristelli, Batty, and Pietronero(2012)}]{cristelli2012zipf}
Cristelli, M.; Batty, M.; and Pietronero, L. 2012.
\newblock There is More than a Power Law in Zipf.
\newblock \emph{Scientific Reports}, 2: 812.

\bibitem[{Datta et~al.(2010)}]{Datta2010}
Datta, A.; et~al. 2010.
\newblock Decentralized online social networks.
\newblock In \emph{Handbook of social network technologies and applications}, 349--378. Springer.

\bibitem[{DeVito, Birnholtz, and Hancock(2017)}]{devito2017platforms}
DeVito, M.~A.; Birnholtz, J.; and Hancock, J.~T. 2017.
\newblock Platforms, People, and Perception: Using Affordances to Understand Self-Presentation on Social Media.
\newblock In \emph{Proceedings of the 2017 ACM Conference on Computer Supported Cooperative Work and Social Computing}, CSCW '17, 740--754. New York, NY, USA: Association for Computing Machinery.
\newblock ISBN 9781450343350.

\bibitem[{Doctorow(2023)}]{doctorow2023social}
Doctorow, C. 2023.
\newblock Social Quitting.
\newblock \emph{Locus}, 90(1): 29.
\newblock Commentary.

\bibitem[{{FORCE11}(2020)}]{fair}
{FORCE11}. 2020.
\newblock The FAIR Data principles.
\newblock \url{https://force11.org/info/the-fair-data-principles/}.

\bibitem[{Gebru et~al.(2021)Gebru, Morgenstern, Vecchione, Vaughan, Wallach, Iii, and Crawford}]{gebru2021datasheets}
Gebru, T.; Morgenstern, J.; Vecchione, B.; Vaughan, J.~W.; Wallach, H.; Iii, H.~D.; and Crawford, K. 2021.
\newblock Datasheets for datasets.
\newblock \emph{Communications of the ACM}, 64(12): 86--92.

\bibitem[{Goel et~al.(2010)Goel, Broder, Gabrilovich, and Pang}]{goel2010anatomy}
Goel, S.; Broder, A.; Gabrilovich, E.; and Pang, B. 2010.
\newblock Anatomy of the Long Tail: Ordinary People with Extraordinary Tastes.
\newblock In \emph{Proceedings of the Third ACM International Conference on Web Search and Data Mining}, WSDM ’10, 201–210. New York, NY, USA: Association for Computing Machinery.
\newblock ISBN 9781605588896.

\bibitem[{Green et~al.(2024)Green, McCabe, Shugars, Chwe, Horgan, Cao, and Lazer}]{green2024curation}
Green, J.; McCabe, S.; Shugars, S.; Chwe, H.; Horgan, L.; Cao, S.; and Lazer, D. 2024.
\newblock Curation Bubbles.
\newblock \emph{American Political Science Review}.
\newblock Forthcoming.

\bibitem[{He et~al.(2023)He, Zia, Castro, Raman, Sastry, and Tyson}]{he2023flocking}
He, J.; Zia, H.~B.; Castro, I.; Raman, A.; Sastry, N.; and Tyson, G. 2023.
\newblock Flocking to Mastodon: Tracking the Great Twitter Migration.
\newblock In \emph{Proceedings of the 2023 ACM on Internet Measurement Conference}, IMC '23, 111–123. New York, NY, USA: Association for Computing Machinery.
\newblock ISBN 9798400703829.

\bibitem[{Hogg et~al.(2024)Hogg, DiResta, Fukuyama, Reisman, Keller, Ovadya, Thorburn, Stray, and Mathur}]{hogg2024shaping}
Hogg, L.; DiResta, R.; Fukuyama, F.; Reisman, R.; Keller, D.; Ovadya, A.; Thorburn, L.; Stray, J.; and Mathur, S. 2024.
\newblock Shaping the {Future} of {Social} {Media} with {Middleware}.
\newblock Technical report, CoRR.
\newblock ArXiv:2412.10283 [cs].

\bibitem[{Hou and Shiau(2020)}]{hou2020understanding}
Hou, A.; and Shiau, W.-L. 2020.
\newblock Understanding Facebook to Instagram migration: a push-pull migration model perspective.
\newblock \emph{Information Technology \& People}, 33(1): 272--295.

\bibitem[{Jeong et~al.(2024)Jeong, Sheth, Tahir, Alatawi, Bernard, and Liu}]{jeong2024exploring}
Jeong, U.; Sheth, P.; Tahir, A.; Alatawi, F.; Bernard, H.~R.; and Liu, H. 2024.
\newblock Exploring Platform Migration Patterns between Twitter and Mastodon: A User Behavior Study.
\newblock \emph{Proceedings of the International AAAI Conference on Web and Social Media}, 18(1): 738--750.

\bibitem[{Kates et~al.(2021)Kates, Tucker, Nagler, and Bonneau}]{kates2021times}
Kates, S.; Tucker, J.; Nagler, J.; and Bonneau, R. 2021.
\newblock The Times They Are Rarely A-Changin’: Circadian Regularities in Social Media Use.
\newblock \emph{Journal of Quantitative Description: Digital Media}, 1.

\bibitem[{Kleppmann et~al.(2024)Kleppmann, Frazee, Gold, Graber, Holmgren, Ivy, Johnson, Newbold, and Volpert}]{kleppmann2024bluesky}
Kleppmann, M.; Frazee, P.; Gold, J.; Graber, J.; Holmgren, D.; Ivy, D.; Johnson, J.; Newbold, B.; and Volpert, J. 2024.
\newblock Bluesky and the AT Protocol: Usable Decentralized Social Media.
\newblock In \emph{Proceedings of the ACM Conext-2024 Workshop on the Decentralization of the Internet}, DIN '24, 1–7. New York, NY, USA: Association for Computing Machinery.
\newblock ISBN 9798400712524.

\bibitem[{La~Cava, Aiello, and Tagarelli(2023)}]{lacava2023drivers}
La~Cava, L.; Aiello, L.~M.; and Tagarelli, A. 2023.
\newblock Drivers of social influence in the {Twitter} migration to {Mastodon}.
\newblock \emph{Scientific Reports}, 13(1): 21626.

\bibitem[{La~Cava, Greco, and Tagarelli(2021)}]{lacava2021}
La~Cava, L.; Greco, S.; and Tagarelli, A. 2021.
\newblock Understanding the growth of the Fediverse through the lens of Mastodon.
\newblock \emph{Applied Network Science}, 6(1): 1--15.

\bibitem[{Lazer et~al.(2018)Lazer, Baum, Benkler, Berinsky, Greenhill, Menczer, Metzger, Nyhan, Pennycook, Rothschild, Schudson, Sloman, Sunstein, Thorson, Watts, and Zittrain}]{lazer2018science}
Lazer, D. M.~J.; Baum, M.~A.; Benkler, Y.; Berinsky, A.~J.; Greenhill, K.~M.; Menczer, F.; Metzger, M.~J.; Nyhan, B.; Pennycook, G.; Rothschild, D.; Schudson, M.; Sloman, S.~A.; Sunstein, C.~R.; Thorson, E.~A.; Watts, D.~J.; and Zittrain, J.~L. 2018.
\newblock The science of fake news.
\newblock \emph{Science}, 359(6380): 1094--1096.

\bibitem[{Lemmer-Webber et~al.(2018)Lemmer-Webber, Tallon, Shepherd, Guy, and Prodromou}]{lemmerwebber2018activitypub}
Lemmer-Webber, C.; Tallon, J.; Shepherd, E.; Guy, A.; and Prodromou, E. 2018.
\newblock {ActivityPub}.
\newblock {W3C} recommendation, World Wide Web Consortium, Wakefield, MA 01880, USA.

\bibitem[{Lin et~al.(2023)Lin, Lasser, Lewandowsky, Cole, Gully, Rand, and Pennycook}]{lin2023high}
Lin, H.; Lasser, J.; Lewandowsky, S.; Cole, R.; Gully, A.; Rand, D.~G.; and Pennycook, G. 2023.
\newblock High level of correspondence across different news domain quality rating sets.
\newblock \emph{PNAS Nexus}, 2(9).

\bibitem[{Mendelsohn et~al.(2023)Mendelsohn, Le~Bras, Choi, and Sap}]{mendelsohn2023dogwhistles}
Mendelsohn, J.; Le~Bras, R.; Choi, Y.; and Sap, M. 2023.
\newblock From {Dogwhistles} to {Bullhorns}: {Unveiling} {Coded} {Rhetoric} with {Language} {Models}.
\newblock In Rogers, A.; Boyd-Graber, J.; and Okazaki, N., eds., \emph{Proceedings of the 61st {Annual} {Meeting} of the {Association} for {Computational} {Linguistics} ({Volume} 1: {Long} {Papers})}, 15162--15180. Toronto, Canada: Association for Computational Linguistics.

\bibitem[{Monroe, Colaresi, and Quinn(2008)}]{Monroe2017}
Monroe, B.; Colaresi, M.; and Quinn, K. 2008.
\newblock Fightin’ Words: Lexical Feature Selection and Evaluation for Identifying the Content of Political Conflict.
\newblock \emph{Political Analysis}, 16(4): 372--403.

\bibitem[{Murtfeldt et~al.(2024)Murtfeldt, Alterman, Kahveci, and West}]{murtfeldt2024rip}
Murtfeldt, R.; Alterman, N.; Kahveci, I.; and West, J.~D. 2024.
\newblock {RIP} {Twitter} {API}: {A} eulogy to its vast research contributions.
\newblock Technical report, CoRR.
\newblock ArXiv:2404.07340 [cs].

\bibitem[{{NewsGuard}(2024)}]{newsguard2024}
{NewsGuard}. 2024.
\newblock NewsGuard Rating Process and Criteria.
\newblock Accessed: January 10, 2024.

\bibitem[{Quelle and Bovet(2024)}]{quelle2024}
Quelle, D.; and Bovet, A. 2024.
\newblock Bluesky: Network Topology, Polarisation, and Algorithmic Curation.
\newblock Technical report, CoRR.
\newblock ArXiv:2405.17571 [cs].

\bibitem[{Raman et~al.(2019)Raman, Joglekar, de~Cristofaro, Sastry, and Tyson}]{raman2019}
Raman, A.; Joglekar, S.; de~Cristofaro, E.; Sastry, N.; and Tyson, G. 2019.
\newblock Challenges in the decentralised web: The Mastodon case.
\newblock In \emph{Proceedings of the ACM SIGCOMM Internet Measurement Conference (IMC)}, 217--229. ACM.

\bibitem[{Resnick, Ovadya, and Gilchrist(2023)}]{resnick2023iffy}
Resnick, P.; Ovadya, A.; and Gilchrist, G. 2023.
\newblock Iffy Quotient: A Platform Health Metric for Misinformation.
\newblock White Paper~v3, University of Michigan Center for Social Media Responsibility, Ann Arbor, MI, USA.

\bibitem[{Reuben et~al.(2024)Reuben, Friedland, Puzis, and Grinberg}]{reuben2024leveraging}
Reuben, M.; Friedland, L.; Puzis, R.; and Grinberg, N. 2024.
\newblock Leveraging {Exposure} {Networks} for {Detecting} {Fake} {News} {Sources}.
\newblock In \emph{Proceedings of the 30th {ACM} {SIGKDD} {Conference} on {Knowledge} {Discovery} and {Data} {Mining}}, {KDD} '24, 5635--5646. New York, NY, USA: Association for Computing Machinery.
\newblock ISBN 9798400704901.

\bibitem[{Ribeiro and Faloutsos(2015)}]{ribeiro2015modeling}
Ribeiro, B.; and Faloutsos, C. 2015.
\newblock Modeling {Website} {Popularity} {Competition} in the {Attention}-{Activity} {Marketplace}.
\newblock In \emph{Proceedings of the {Eighth} {ACM} {International} {Conference} on {Web} {Search} and {Data} {Mining}}, {WSDM} '15, 389--398. New York, NY, USA: Association for Computing Machinery.
\newblock ISBN 9781450333177.

\bibitem[{Schneider(2019)}]{schneider2019decentralization}
Schneider, N. 2019.
\newblock Decentralization: an incomplete ambition.
\newblock \emph{Journal of Cultural Economy}, 12(4): 265--285.

\bibitem[{Shao et~al.(2016)Shao, Ciampaglia, Flammini, and Menczer}]{Shao2016}
Shao, C.; Ciampaglia, G.~L.; Flammini, A.; and Menczer, F. 2016.
\newblock Hoaxy: A Platform for Tracking Online Misinformation.
\newblock In \emph{Proceedings of the 25\textsuperscript{th} International Conference Companion on World Wide Web}, WWW '16 Companion, 745--750. Republic and Canton of Geneva, Switzerland: International World Wide Web Conferences Steering Committee.
\newblock ISBN 978-1-4503-4144-8.

\bibitem[{Shao et~al.(2018{\natexlab{a}})Shao, Ciampaglia, Varol, Yang, Flammini, and Menczer}]{shao2018spread}
Shao, C.; Ciampaglia, G.~L.; Varol, O.; Yang, K.; Flammini, A.; and Menczer, F. 2018{\natexlab{a}}.
\newblock The spread of low-credibility content by social bots.
\newblock \emph{Nature Communications}, 9(1): 4787.

\bibitem[{Shao et~al.(2018{\natexlab{b}})Shao, Hui, Wang, Jiang, Flammini, Menczer, and Ciampaglia}]{shao2018anatomy}
Shao, C.; Hui, P.-M.; Wang, L.; Jiang, X.; Flammini, A.; Menczer, F.; and Ciampaglia, G.~L. 2018{\natexlab{b}}.
\newblock Anatomy of an online misinformation network.
\newblock \emph{PLOS ONE}, 13(4): 1--23.

\bibitem[{Wähner et~al.(2024)Wähner, Deubel, Breuer, and Weller}]{waehner2024dont}
Wähner, M.; Deubel, A.; Breuer, J.; and Weller, K. 2024.
\newblock “{Don}’t research us”—{How} {Mastodon} instance rules connect to research ethics.
\newblock \emph{Publizistik}, 69(3): 357--380.

\bibitem[{Zhang et~al.(2024)Zhang, Zhao, Wang, Johnston, Chalhoub, Ross, Liu, Tinsman, Zhao, Van~Kleek, and Shadbolt}]{Zhang2024}
Zhang, Z.; Zhao, J.; Wang, G.; Johnston, S.-K.; Chalhoub, G.; Ross, T.; Liu, D.; Tinsman, C.; Zhao, R.; Van~Kleek, M.; and Shadbolt, N. 2024.
\newblock Trouble in Paradise? Understanding Mastodon Admin's Motivations, Experiences, and Challenges Running Decentralised Social Media.
\newblock \emph{Proc. ACM Hum.-Comput. Interact.}, 8(CSCW2).

\end{thebibliography}

\section{Paper Checklist -- Ethics Guidelines}
\begin{enumerate}

\item For most authors...
\begin{enumerate}
    \item  Would answering this research question advance science without violating social contracts, such as violating privacy norms, perpetuating unfair profiling, exacerbating the socio-economic divide, or implying disrespect to societies or cultures?
    \answerYes{Yes}
  \item Do your main claims in the abstract and introduction accurately reflect the paper's contributions and scope?
    \answerYes{Yes}
   \item Do you clarify how the proposed methodological approach is appropriate for the claims made? 
    \answerYes{Yes}
   \item Do you clarify what are possible artifacts in the data used, given population-specific distributions?
    \answerYes{Yes, this is mentioned in the Limitations and Ethical Considerations section.}
  \item Did you describe the limitations of your work?
    \answerYes{Yes, in the Limitations and Ethical Considerations section.}
  \item Did you discuss any potential negative societal impacts of your work?
    \answerYes{Yes, in the Limitations and Ethical Considerations section.}
      \item Did you discuss any potential misuse of your work?
    \answerYes{Yes, in the Limitations and Ethical Considerations section.}
    \item Did you describe steps taken to prevent or mitigate potential negative outcomes of the research, such as data and model documentation, data anonymization, responsible release, access control, and the reproducibility of findings?
    \answerYes{Yes, in the Limitations and Ethical Considerations.}
  \item Have you read the ethics review guidelines and ensured that your paper conforms to them?
    \answerYes{Yes.}
\end{enumerate}

\item Additionally, if your study involves hypotheses testing...
\begin{enumerate}
  \item Did you clearly state the assumptions underlying all theoretical results?
    \answerNA{NA}
  \item Have you provided justifications for all theoretical results?
    \answerNA{NA}
  \item Did you discuss competing hypotheses or theories that might challenge or complement your theoretical results?
    \answerNA{NA}
  \item Have you considered alternative mechanisms or explanations that might account for the same outcomes observed in your study?
    \answerNA{NA}
  \item Did you address potential biases or limitations in your theoretical framework?
    \answerNA{NA}
  \item Have you related your theoretical results to the existing literature in social science?
    \answerNA{NA}
  \item Did you discuss the implications of your theoretical results for policy, practice, or further research in the social science domain?
    \answerNA{NA}
\end{enumerate}

\item Additionally, if you are including theoretical proofs...
\begin{enumerate}
  \item Did you state the full set of assumptions of all theoretical results?
    \answerNA{NA}
	\item Did you include complete proofs of all theoretical results?
    \answerNA{NA}
\end{enumerate}

\item Additionally, if you ran machine learning experiments...
\begin{enumerate}
  \item Did you include the code, data, and instructions needed to reproduce the main experimental results (either in the supplemental material or as a URL)?
    \answerNA{NA}
  \item Did you specify all the training details (e.g., data splits, hyperparameters, how they were chosen)?
    \answerNA{NA}
     \item Did you report error bars (e.g., with respect to the random seed after running experiments multiple times)?
    \answerNA{NA}
	\item Did you include the total amount of compute and the type of resources used (e.g., type of GPUs, internal cluster, or cloud provider)?
    \answerNA{NA}
     \item Do you justify how the proposed evaluation is sufficient and appropriate to the claims made? 
    \answerNA{NA}
     \item Do you discuss what is ``the cost`` of misclassification and fault (in)tolerance?
    \answerNA{NA}
  
\end{enumerate}

\item Additionally, if you are using existing assets (e.g., code, data, models) or curating/releasing new assets, \textbf{without compromising anonymity}...
\begin{enumerate}
  \item If your work uses existing assets, did you cite the creators?
    \answerYes{Yes}
  \item Did you mention the license of the assets?
    \answerNA{NA}
  \item Did you include any new assets in the supplemental material or as a URL?
    \answerNA{NA}
  \item Did you discuss whether and how consent was obtained from people whose data you're using/curating?
    \answerNo{No, because we are analyzing publicly available Bluesky data collected with a social listening tool and so obtaining consent for this collection is not practical.}
  \item Did you discuss whether the data you are using/curating contains personally identifiable information or offensive content?
    \answerYes{Yes}
\item If you are curating or releasing new datasets, did you discuss how you intend to make your datasets FAIR (see \citet{fair})?
    \answerNA{NA}
\item If you are curating or releasing new datasets, did you create a Datasheet for the Dataset (see \citet{gebru2021datasheets})? 
    \answerNA{NA}
\end{enumerate}

\item Additionally, if you used crowdsourcing or conducted research with human subjects, \textbf{without compromising anonymity}...
\begin{enumerate}
  \item Did you include the full text of instructions given to participants and screenshots?
    \answerNA{NA}
  \item Did you describe any potential participant risks, with mentions of Institutional Review Board (IRB) approvals?
    \answerNA{NA}
  \item Did you include the estimated hourly wage paid to participants and the total amount spent on participant compensation?
    \answerNA{NA}
   \item Did you discuss how data is stored, shared, and deidentified?
    \answerNA{NA}
\end{enumerate}

\end{enumerate}

\end{document}